\documentclass[%
 reprint,
 amsmath,amssymb,
 aps,
prb,
]{revtex4-1}

\usepackage{amsmath}
\usepackage{hyperref}
\usepackage{graphicx} 
\usepackage{dcolumn}  
\usepackage{bm}       
\usepackage{graphicx} 
\usepackage{float}    
\usepackage{wrapfig}  
\usepackage{comment}
\usepackage[T2A]{fontenc}
\usepackage[utf8]{inputenc}
\usepackage[russian,english]{babel}
\usepackage{caption}
\usepackage{subcaption}
\usepackage{blindtext}

\DeclareRobustCommand{\[}{\begin{gather*}}
\DeclareRobustCommand{\]}{\end{gather*}}

\begin{document}
\title{Plasmonic drag photocurrent in graphene at extreme nonlocality}
\author{Vladimir Silkin}
\author{Dmitry Svintsov}
\affiliation{Laboratory of 2d Materials for Optoelectronics, Moscow Institute of Physics and Technology, Dolgoprudny 141700, Russia}

\begin{abstract}
It is commonly assumed that photocurrent in two-dimensional systems with centrosymmetric lattice is generated at structural inhomogenities, such as p-n junctions. Here, we study an alternative mechanism of photocurrent generation associated with inhomogenity of the driving electromagnetic field, termed as 'plasmonic drag'. It is associated with direct momentum transfer from field to conduction electrons, and can be characterized by a non-local non-linear conductivity $\sigma^{(2)}({\bf q},\omega)$. By constructing a classical kinetic model of non-linear conductivity with full account of non-locality, we show that it is resonantly enhanced for wave phase velocity coinciding with electron Fermi velocity. The enhancement is interpreted as phase locking between electrons and the wave. We discuss a possible experiment where non-uniform field is created by a propagating graphene plasmon, and find an upper limit of the current responsivity vs plasmon velocity. This limit is set by a competition between resonantly growing $\sigma^{(2)}({\bf q},\omega)$ and diverging kinetic energy of electrons as the wave velocity approaches Fermi velocity.
\end{abstract}

\maketitle

\section{\label{sec:level1}Introduction}

It is commonly believed that photocurrent generation in two-dimensional systems without lattice inversion asymmetry occurs at the structural inhomogenities, such as p-n junctions~\cite{Song2011,Muravev2012a,Tielrooij2015} and contacts with metals~\cite{Cai2014,Echtermeyer2014}. At the same time, the inversion asymmetry of driving electromagnetic field itself can lead to emergence of photocurrent. The physics beyond such photocurrent is the direct transfer of electromagnetic field momentum to the electrons, termed in literature as photon drag~\cite{Photon-drag-Ge}, or dynamic Hall effect~\cite{Karch_dynamic_Hall}. As the optoelectronic structures become deeply sub-wavelength, their electromagnetic response is dominated by near-fields. The mechanism of current generation by field with finite momentum is thus termed as plasmonic drag~\cite{Photon_drag_by_plasmons,Noginova_Plasmonic_drag,Popov_plasmonic_drag}, implying that momentum is transferred via excitation of two-dimensional plasmons.

\begin{figure}[h!]
    \includegraphics[width=0.9\linewidth]{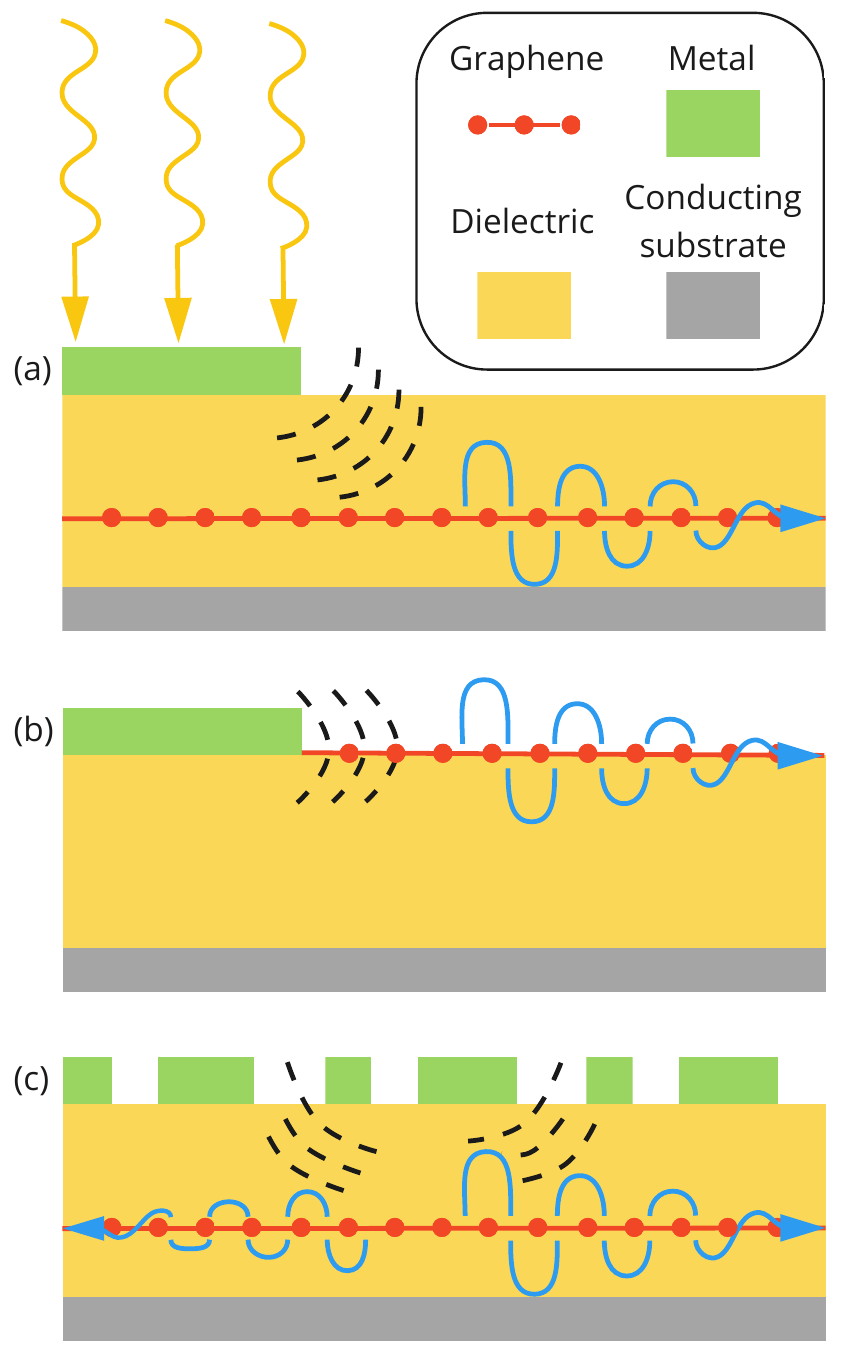}
    \caption{Possible structures where plasmonic drag can be observed upon diffraction of incident plane electromagnetic wave: (a) graphene under the edge of a metal gate; (b) graphene contacted by metal (c) graphene under the grating lacking inversion symmetry}
    \label{fig:Structures}
\end{figure}

Plasmonic drag photocurrent is ubiquitous to systems with uniform channel and asymmetric electromagnetic environment. Experimental examples of such environments are gate edges~\cite{Song2011}, metal contacts, and grating couplers without inversion symmetry~\cite{Olbrich2016,Popov2015} (Fig.~\ref{fig:Structures}). Plasmonic drag can occur upon tilted illumination of fully symmetric structure with identical contacts, or illumination of one of its contacts which acts as photon-to-plasmon coupler~\cite{Nikulin_EdgeDiffraction}. Compared to photovoltaic and photo-thermoelectric effects, the plasmonic drag does not need channel doping non-uniformity. We may therefore speculate that it is the most omnipresent mechanism of photocurrent generation in two dimensions~\footnote{The plasmonic drag effect is conceptually equivalent to the distributed resistive self-mixing~\cite{Sakowicz2011} and Dyakonov-Shur~\cite{Sakowicz2011} rectification. This can be seen by presenting the photocurrent via linear-response electric field and its gradients, all three mechanisms will be described by identical expressions.}.

As the size of the optoelectronic devices $L$ is shrinking, the characteristic wave number $q$ of electromagnetic near fields scales up as $L^{-1}$. Down-scaling of the gate-channel separation also leads to slow-down of plasmons~\cite{Chaplik1972} and increase in characteristic wave vector at given frequency. At very large wave vectors, the current-field relations become non-local and reflect the dynamics of individual electrons. While linear non-local response of two-dimensional systems is quite well studied~\cite{Heitmann_FermiPressure,bandurin2021cyclotron,Lundeberg2017c}, the non-linear processes of photocurrent generation at strong non-locality are almost unexplored. Indeed, previous studies of plasmonic drag were limited either to expansions linear in field wave vector~\cite{Olbrich2016}, or to the hydrodynamic description of electron systems~\cite{Popov_plasmonic_drag}. The latter is valid only at small frequencies compared to that of carrier-carrier collisions, and at long wavelengths much exceeding the mean free path~\cite{Svintsov2018c}.

In this Letter, we theoretically explore the limits of plasmonic drag in graphene at arbitrarily strong non-locality, i.e. at arbitrarily large field wave vector $q$. Our choice of graphene is dictated by recent observations of ultra-confined acoustic graphene plasmons~\cite{Lundeberg2017c,Iranzo2018,Epstein2020,Bylinkin2019} which spectrum is governed by non-locality, and their potential for photocurrent harvesting~\cite{Bandurin2018d}. We derive the non-local non-linear conductivity of graphene $\sigma^{(2)({\bf q},\omega)}$ being the proportionality coefficient between photocurrent density and squared ac electric field, ${\bf j}^{(2)} = {\bf n}_{\bf q} \sigma^{(2)({\bf q},\omega)} E^2$. It possesses a square-root singularity at phase velocity $\omega/q$ approaching the Fermi velocity of 2d electrons $v_0$. We interpret this singularity as phase locking between dragged electrons and electromagnetic field. Second, we quantify the electromagnetic energy flux $S_{\rm pl}$ carried by 2d plasmons at large $q$. The ratio of these quantities is nothing but current responsivity measured in photodetetion experiments. The energy flux has a counter-balancing singularity to that in $\sigma^{(2)}$, it appears due to large kinetic energy of charge carriers in electromagnetic wave. As a result, the photocurrent responsivity has a universal maximum order of $0.25 e/E_F$ achieved at $\omega \approx 1.4 q v_0$.

The starting point for evaluation of photocurrent in non-uniform field ${\bf E}({\bf r},t)$ is the classical kinetic equation for electron distribution function $f({\bf r},{\bf p},t)$:
\begin{equation}
    \frac{\partial f}{\partial t}+ {\bf v_p} \frac{\partial f}{\partial {\bf r}} - e {\bf E}({\bf r},t) \frac{\partial f}{\partial {\bf p}} = -\frac{f-f_{0}}{\tau_p},
\end{equation}
where ${\bf v}_{\bf p}$ is the electron velocity, $\tau_{p}$ is the momentum relaxation time, and $f_{0}$ is the equilibrium (Fermi) distribution function. We specify the field in the form of running wave ${\bf E}({\bf r},t) = {\bf E}_0 \: e^{i({\bf q} {\bf r} - \omega t)} + {\rm h.c.}$, where ${\rm h.c.}$ stands for complex conjugate, and limit ourselves to purely longitudinal fields ${\bf E}_0 \parallel {\bf q}$. We adopt successive approximations to distribution function in powers of the field, $f=f_{0} + f_1 
+f_2 
$, where $f_n \propto E_0^{n}$.

\begin{figure}[h!]
    \centering
    \includegraphics[width=0.9\linewidth]{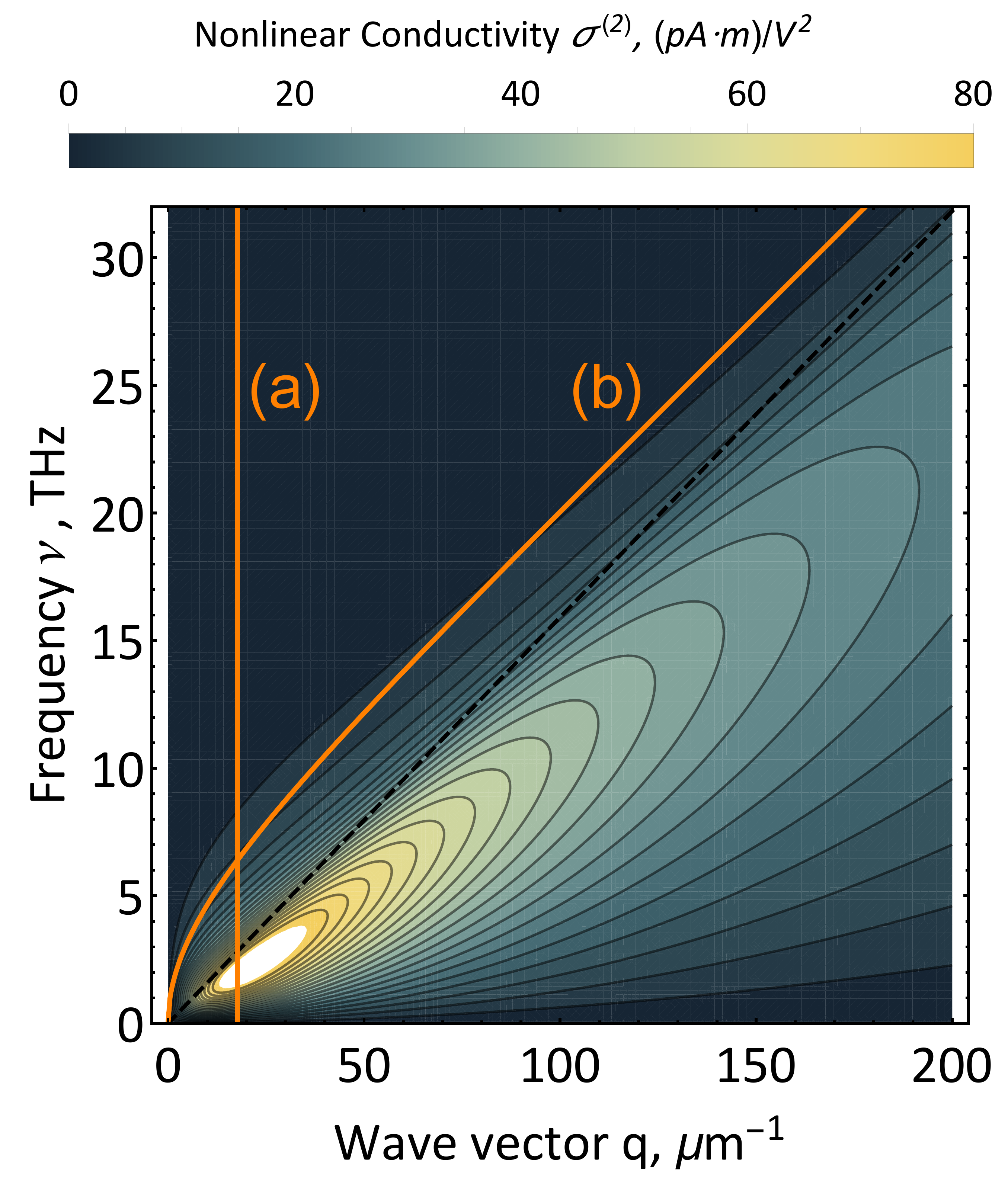}
    \caption{Color map of non-linear conductivity $\sigma^{(2)}(q,\omega)$ vs frequency $\omega/2\pi$ and wave vector $q$. Vertical line (a) corresponds to constant wave vector $q=q_F$; line (b) shows the dispersion of graphene plasmons. Momentum relaxation time $\tau_p = 10^{-13}$ s, carrier density $n = 10^{10}$ cm$^{-2}$.}
    \label{fig:Sigma-2-contour}
\end{figure}


The quantity of interest for calculation of dc photocurrent is the to time- and space-averaged  second-order distribution function $\langle f_2 \rangle_{t,x}$. Performing lengthy yet common calculations, we find it in the form
\begin{equation}
    \label{f2}
    {\langle f_2 \rangle}_{t,x} =\frac{e^2 E_{0}^2}{2}\frac{\partial}{\partial {\bf p}} \Bigg\{ \frac{\partial f_{0} / \partial {\bf p}}{(\omega - {\bf q v_p})^2+{\tau_p}^{-2}} \Bigg\}.
\end{equation}

The rectified current is obtained by integration over momentum space ${\bf j}^{(2)} = g \sum\limits_{\bf p}{{\langle f_2 \rangle}_{t,x} }$, where $g=4$ is the electron degeneracy factor in graphene. Apparently, the rectified current is proportional to $E_0^2$ and directed along ${\bf n_q} = {\bf q}/q$. Introducing the nonlinear conductivity ${\bf j}^{(2)} = \sigma^{(2)}(q,\omega) E_0^2 {\bf n_q}$, we find (see Appendix A for details):
\begin{equation}\label{Sigma-2}
    \sigma^{(2)} (q,\omega) =  -\frac{g e^3}{2 \pi \hbar^2 v_0}
    \frac{f_{0}(0)}{q^2}
    \left ( 
    \frac{2\,\widetilde{v}_{ph}^2-1}{2\,\sqrt{\widetilde{v}_{ph}^2-1}}-\widetilde{v}_{ph},
    \right )
\end{equation}
where $\widetilde{v}_{ph} = (\omega + i \tau_p^{-1})/(q v_0)$  is the dimensionless phase velocity, and $f_{0}(0)$ is the equilibrium electron distribution function evaluated at zero energy. The photocurrent due to drag of holes is additive with opposite sign, thus account of both sorts of carriers amounts to replacement $f_{0}(0) \rightarrow 2f_{0}(0) - 1$.

The most remarkable property of nonlinear conductivity (\ref{Sigma-2}) is the presence of square-root singularity as the wave phase velocity $\omega/q$ approaches the electron Fermi velocity from either side, as shown in Fig.~\ref{fig:Sigma-2-contour}. A detailed inspection shows that singular contribution to current comes from electrons moving in phase with the wave, i.e. at angles $\cos \theta \approx \omega/(q v_0)$. We may say that these electrons are trapped in the minima of harmonic potential induced by the wave, and move synchronously with the wave velocity.

\begin{figure}[h!]
    \centering
    \includegraphics[width=0.9\linewidth]{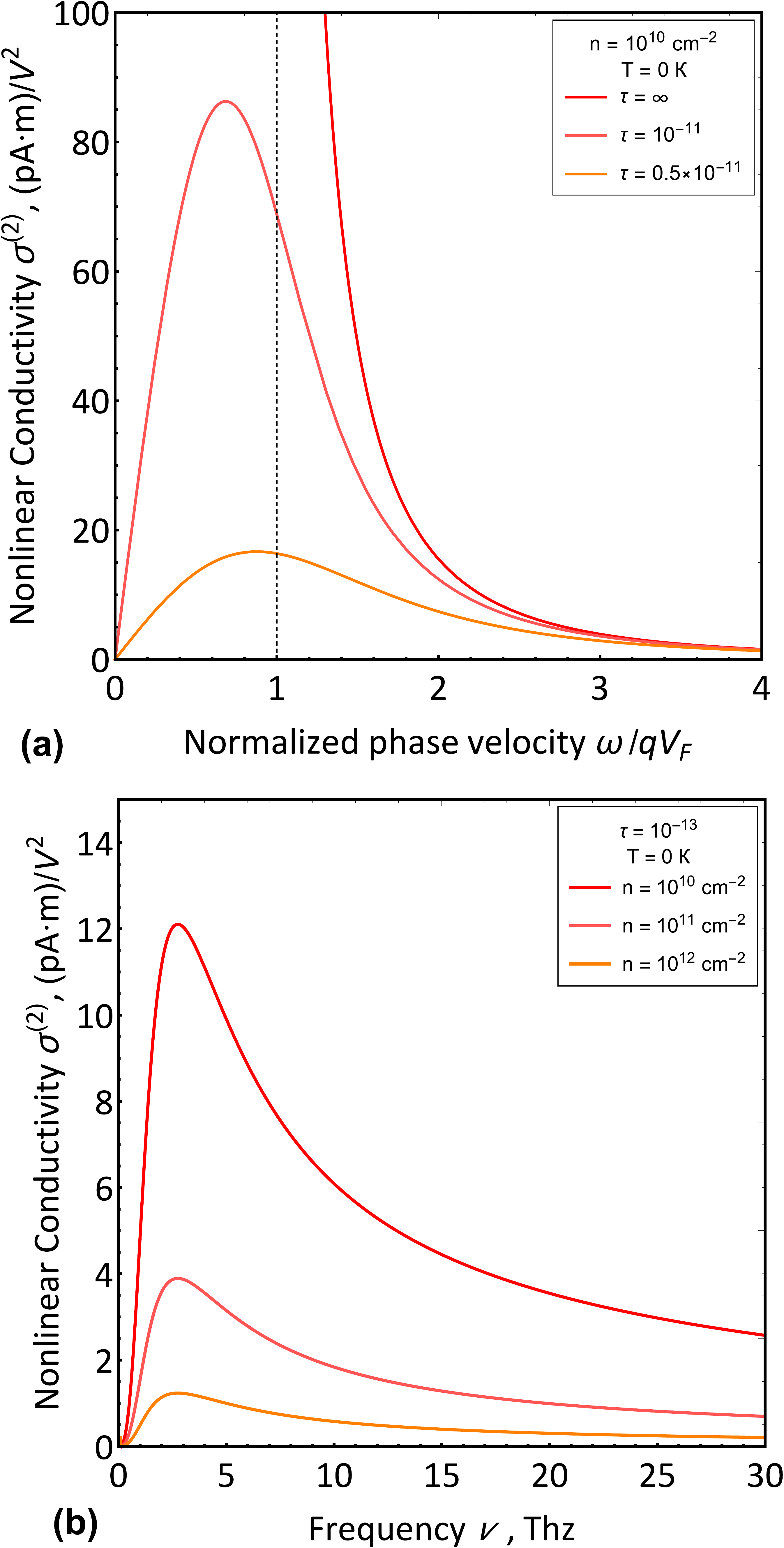}
    \caption{ Dependence of the nonlinear conductivity $\sigma^{(2)}$ on normalized phase velocity at (a) constant wave vector $q=q_F = 20$ $\mu$m$^{-1}$, corresponding to the line cut (a) in Fig.~\ref{fig:Sigma-2-contour} (b) $\omega$ and $q$ bound by plasmon dispersion at various carrier densities $n=10^{10}$, $10^{11}$ and $10^{12}$ cm$^{-2}$. Momentum relaxation time $\tau_p = 10^{-13}$ s.}
    \label{fig:Sigma-2-cuts}
\end{figure}

The effect of finite scattering rate $\tau_p^{-1}$ on nonlinear conductivity is highly non-trivial. It depends critically on the ratio of phase velocity $\omega/q$ and Fermi velocity $v_0$, which is illustrated in Fig.~\ref{fig:Sigma-2-cuts} (a). For slow waves with $\omega/qv_0 < 1$, the non-linear conductivity $\sigma^{(2)}$ is approximately proportional to relaxation time. This result is interpreted as follows: the amount of momentum transferred from field to electrons is proportional to the rate of Landau damping. Being a collisionless process, it does not depend on $\tau_p$. The current established at given momentum transfer rate (i.e. at given force density) is inversely proportional to electron scattering rate. 

The situation for fast waves, $\omega/qv_0 > 1$, is different. Such waves can not induce intraband Landau damping by the virtue of momentum conservation, and finite scattering rate is required to soften the momentum constraint. At the same time, scattering acts to dissipate the generated current. As a result, $\sigma^{(2)}$ becomes independent of scattering time for relatively fast phase velocities $\omega/qv_0 \gg 1$. This rule breaks down as we approach the singularity at $\omega = qv_0$ which is softened by scattering at both sides.


The above description of rectified photocurrent was performed at given field ${\bf E}_0 ({\bf q},\omega)$ in the plane of graphene. Determination of this field requires a solution of linear diffraction problem for a wave scattered by graphene and its electromagnetic environment (gratings~\cite{Fateev2017}, tips~\cite{McLeod_PRB_NearField}, contacts~\cite{Nikulin_EdgeDiffraction}, etc.). According to the results of analytical and numerical studies, the field in the plane of graphene is dominated by 2d plasmons. Their wave vector is bound to frequency via dispersion relation~\cite{Ryzhii2006c}:
\begin{equation}
\label{Eq-dispersion}
    \omega_{\rm pl}(q)= q v_{0} \frac{1+ g \alpha q_F/q }{\sqrt{1+2 g \alpha q_{F}/q}},
\end{equation}
where $\alpha = e^2/(\kappa\hbar v_0)$ is the coupling constant in graphene, $\kappa$ is the background dielectric constant, and $q_F$ is the electron Fermi wave vector.

A distinctive property of dispersion (\ref{Eq-dispersion}) lies in complete absence of Landau damping. This is guaranteed by the phase velocity of graphene plasmon that always exceeds Fermi velocity $v_0$. It is thus tempting to see whether the the singularity in non-linear conductivity, Eq.~(\ref{Sigma-2}), can be probed with graphene plasmons. 

To test this, we bind $q$ and $\omega$ via plasmon dispersion and plot $\sigma^{(2)}$ parametrically via $v_{\rm ph}$ (which now cannot fall below unity). The result is shown in Fig.~\ref{fig:Sigma-2-cuts} (b). Once the phase velocity is large, $v_{\rm ph} \gg 1$, the nonlinear conductivity rolls down. Indeed, large phase velocities correspond to initial part of plasmon dispersion curve where the characteristic wave vector (transferred momentum) is small. Once the plasmon phase velocity approaches Fermi velocity, $\sigma^{(2)}$ becomes small again. This fact may look inconsistent with the presence of singularity in $\sigma^{(2)}$, but it is. Indeed, synchronism of wave velocity and Fermi velocity is achieved only at very high frequencies, where inertial electrons do not keep up with rapidly oscillating electric field.

The only way to probe the singular non-linear conductivity is thus to bring the whole plasmon dispersion closer to the singular line $\omega = q v_0$. This can be achieved via reduction of carrier density and/or increase in background dielectric constant. The result is illustrated in Fig.~\ref{fig:Sigma-2-cuts} (b), where reducing $n = k_F^2/\pi$ from $10^{12}$ cm$^{-2}$ to $10^{10}$ cm$^{-2}$ leads to an order of magnitude enhancement of non-linear conductivity. 

\begin{figure}[ht]
        \includegraphics[width=0.9\linewidth]{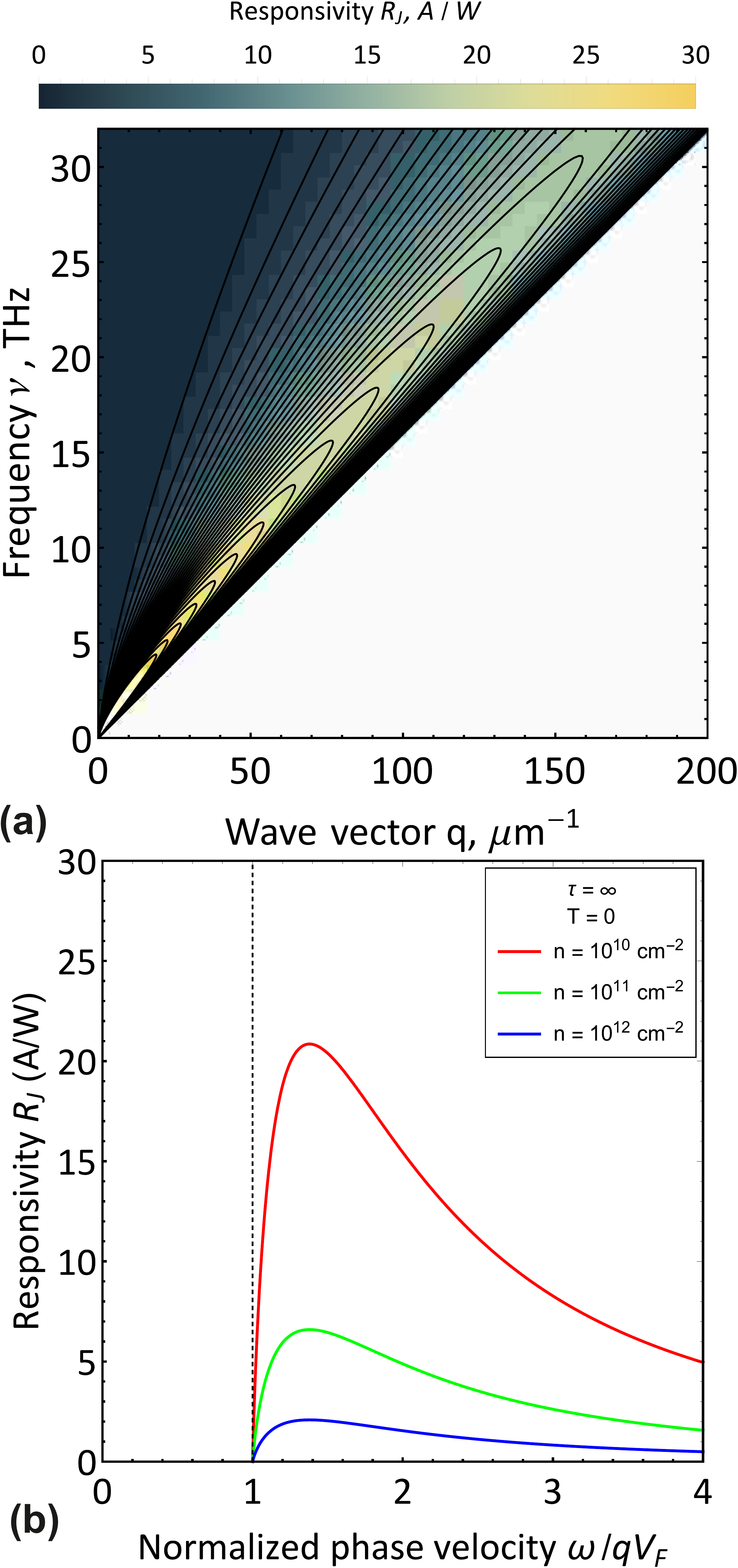}
        \caption{Dependence of the current responsivity $R_J$ on frequency $\omega/2\pi$ and wave vector $q$. Cut lines (a) and (b) correspond to $q = {\rm const}$ and plasmon dispersion relation at carrier density $n=10^{10}$ cm$^{-2}$. Plasmonic drag responsivity evaluated at plasmon dispersion relation at various carrier densities}
   \label{fig:R}  
\end{figure}

Experimental measurement of non-linear conductivity at finite wave vector $q$ is very challenging as the amplitude of electric field $E_0$ in the 2d plane is different from the incident field $E_{\rm inc}$. This difference stems from strong self-consistent field effects which, in fact, are responsible for launching of plasmons. A more common measurable quantity is the photocurrent responsivity $R_{J} = j^{(2)}/S$, where $S$ is the incident power density. Below, we provide an upper bound for plasmonic drag responsivity from energy balance considerations.

Under perfect matching conditions, the power flow of incident electromagnetic wave $S$ is fully transformed into the power flow carried by 2d plasmon $S_{\rm pl}$. A nearly perfect conversion is attainable under proper design of grating couplers. In simpler systems, such as metal edges, the conversion coefficient is well below unity, $S_{\rm pl}/S \ll 1$. Approximating $S_{\rm pl} \approx S$, we obtain a natural upper bound of plasmonic drag responsivity.

The power flow density ${\bf s}$ carried by plasmon in a spatially dispersive 2d system is the sum of electromagnetic and kinetic contributions~\cite{landau2013electrodynamics}:
\begin{equation}
\label{s-density}
    {\bf s} = \frac{c}{8\pi} \left [ {\bf E}\times \textbf{H}^{*}\right ] + \frac{1}{4}
    \frac{\partial {\rm Im} \sigma^{(1)}(q,\omega)}{\partial \textbf{q}}\left ( \textbf{E},\textbf{E}^{*} \right )\delta(z),
\end{equation}
where $\sigma^{(1)}(q,\omega)$ is the linear conductivity of graphene given in the classical limit by
\begin{equation}\label{sigma-1}
    \sigma^{(1)}(q,\omega) = i g \frac{e^2}{\hbar}
    \frac{E_{F}}{2 \pi \hbar}
    \frac{\omega}{q^2 v_{0}^2}
    \Bigg[ \frac{\omega}{\sqrt{\omega^2 - q^2 v_{0}^2}}-1     \Bigg]
\end{equation}

Integrating the flow density (\ref{s-density}) over the vertical coordinate $z$, we obtain the full power flow $S_{\rm pl} = \int{s_x dz}$ in the form
\begin{equation}
   S_{\rm pl} = 
    \left.
    \frac{E_{0}^2}{4}\frac{1}{q}\frac{\partial}{\partial q} \Big[q\,\sigma^{(1)}(q,\omega) \Big] 
    \right|_{{\omega = \omega_{\rm pl}(q)}}.
\end{equation}

It is now apparent that the the power flow carried by a plasmon is diverging as its velocity approaches Fermi velocity. Physically, it comes from very large contribution of carrier kinetic energy to the net energy flow. Formally, it comes from differentiating singular conductivity ${\rm Im}\sigma^{(1)} \propto [\omega^2 - q^2v_0^2]^{-1/2}$; differentiation enhances the strength of singularity.

It is now becoming clear that the current responsivity $R_J = j^{(2)}/S_{\rm pl}$ is bounded from above as the plasmon phase velocity approaches the Fermi velocity. Indeed, a singular growth in $\sigma^{(2)}$ at $\omega/q \rightarrow v_0$ is overwhelmed by a faster growth in $S_{\rm pl}$. This situation is illustrated in Fig.~\ref{fig:R}. Panel (a) shows $R_J$ as a function of independent $q$ and $\omega$, naturally, the maximum of $R_J$ lies above the singular line $\omega = q v_0$. Panel (b) shows the responsivity evaluated at $\omega$ and $q$ bound by dispersion (\ref{Eq-dispersion}); again, this function has a pronounced maximum. 

Instructively, it is possible to present the current responsivity in a universal form at $T=0$:
\begin{equation}
\label{Responsivity-limit}
    R_J = \frac{e}{E_F} f \left ( \widetilde{v}_{ph} \right ),
\end{equation}
where $f \left ( \widetilde{v}_{ph} \right )$ is the dimesnionless function of a single dimensionless parameter, the scaled phase velocity $\widetilde{v}_{ph} = \omega/ q v_0$. The function $f(\widetilde{v}_{ph})$ reaches a maximum value of $0.243$ at $\omega = 1.38 q v_0$. Remarkably, the result does not depend on carrier density and dielectric environment.

\begin{figure}[ht]
    \includegraphics[width=0.9\linewidth]{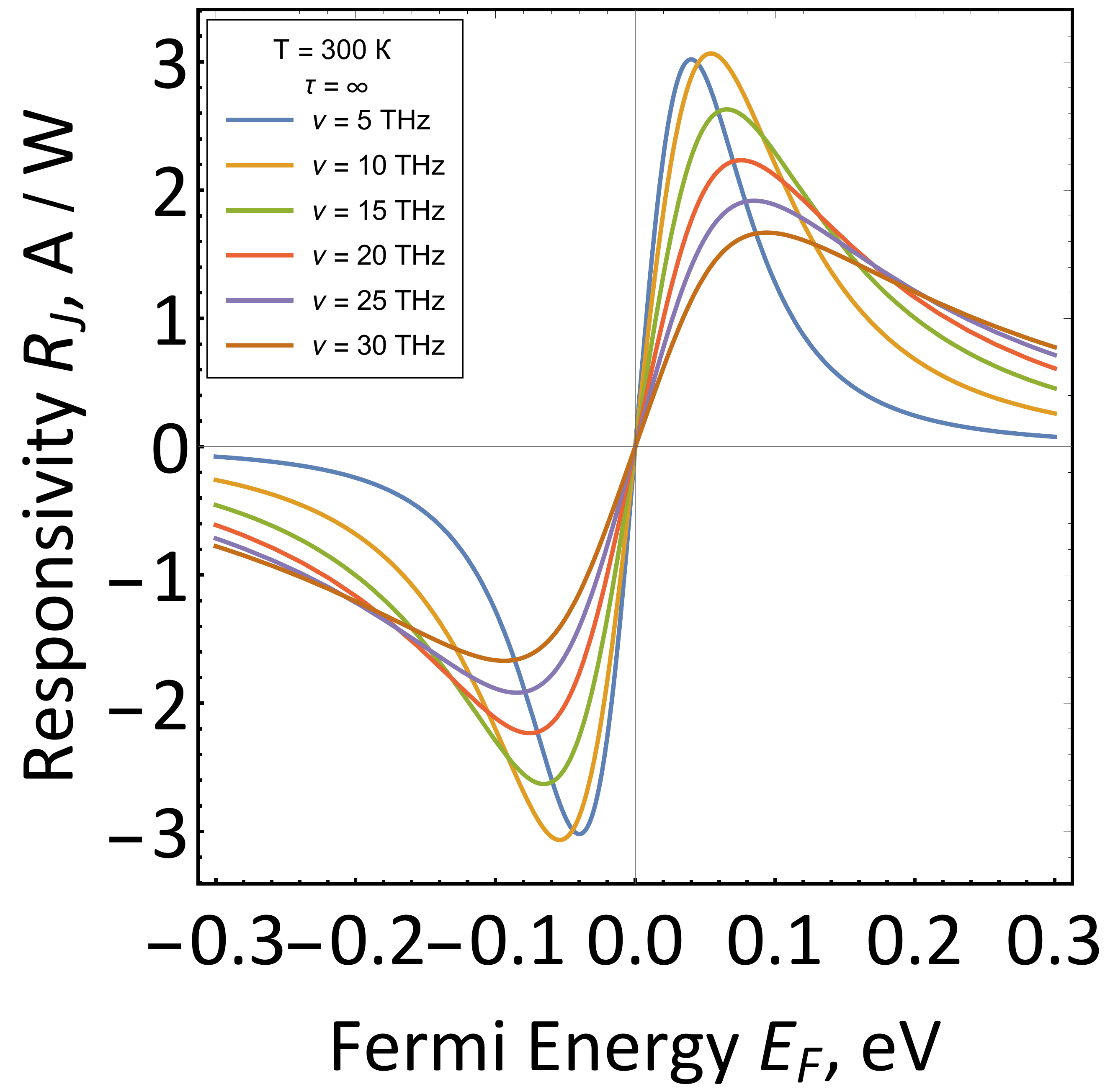}
    \caption{Plasmonic drag responsivity vs Fermi energy evaluated at different frequencies of incoming radiation. The wave vector is obtained from the plasmon dispersion relation. Temperature $T=300$ K.}
    \label{fig:RJ_finiteT}
\end{figure}

The ultimate plasmonic drag responsivity (\ref{Responsivity-limit}) can be compared to that of a perfect photovoltaic cell $R_{\rm pv} < e/\hbar\omega$. Naturally, the drag responsivity is below the photovoltaic limit, as all our calculations were performed in the classical domain $E_F \gg \hbar \omega$. Nevertheless, the maximum responsivities of $2...20$ A/W in Fig.~\ref{fig:R} are large compared to those of typical graphene photodetectors operating in the terahertz~\cite{Bandurin2018,Vicarelli2012a} and infrared~\cite{Badioli2014} frequency ranges. It should be also noted that plasmonic drag mechanism provides fast photoresponse~\cite{Muravev2016}, as the timescale for decay of photocurrent is the momentum relaxation time in the bulk.

All previous calculations carried out at $T=0$ have indicated that plasmonic drag responsivity benefits from low carrier density. At finite temperature, the density is limited by thermal excitation of carriers. The plasmon velocity is sensitive to carrier density, and also cannot become very close to Fermi velocity $v_0$. Account of finite temperature amounts to a simple replacement of 'effective Fermi energy' in expressions for linear conductivity and plasmon dispersion
\begin{equation}
    E_F \rightarrow kT \ln \left( 1+ e^{E_F/kT}\right)\left( 1+ e^{-E_F/kT}\right).
\end{equation}

It is possible to evaluate $R_J$ at finite temperature with full account of thermally excited carriers, which is done in Fig.~\ref{fig:RJ_finiteT} at $T=300$ K. The obtained dependence of plasmonic drag responsivity on Fermi energy is an anti-symmetric function of $E_F$, with a maximum located at $E_F \sim kT$. The functional dependence of $R_J$ on frequency is quite peculiar. Namely, the responsivity grows with increasing frequency at large Fermi energies. This growth is associated with increased plasmon wave vector at higher frequencies, and higher average momentum transferred to an electronic system.

\bibliography{Literature}

\appendix
\begin{widetext}

\section{Сalculation of nonlinear conductivity}

We calculate  $\sigma^{(2)}={\langle j_2 \rangle}_{t,x}/E_0^2$ through the integral of the current. 

\begin{equation}\label{j2}
    {\langle j_2 \rangle}_{t,x} = - \frac{g}{(2 \pi \hbar)^2} e\iint\limits_{-\infty}^{+\infty}v_x {\langle f_2 \rangle}_{t,x} dp_x dp_y
\end{equation}

After Revealing the derivative in ${\langle f_2 \rangle}_{t,x}$ we use polar system of coordinates, where:

\begin{equation}
    \begin{gathered}
         dp_x dp_y = p dp d\phi \\
        \frac{\partial f_0}{\partial p_x} = \frac{\partial f_0}{\partial p} \cos{\phi} \\
        \frac{\partial^2 f_0}{\partial p_x^2} = \cos^2{\phi} \frac{\partial^2 f_0}{\partial p^2} + \frac{\sin^2{\phi}}{p} \frac{\partial f_0}{\partial p} \\
         \frac{\partial v_x}{\partial p_x} = \frac{\sin^2{\phi}}{p} v_0 
    \end{gathered}
\end{equation}
Momentum integrals yield to:
\begin{equation}
    \begin{gathered}
        \int_{0}^{+\infty}\frac{\partial f_0}{\partial p} \,dp= - f_0(0)\\ \int_{0}^{+\infty}\frac{\partial^2 f_0}{\partial p^2} p \,dp= f_0(0)
    \end{gathered}
\end{equation}
so, the $\sigma^{(2)}$ yields:
\begin{equation}\label{sigma2_appendix}
    \boxed{\sigma^{(2)}  = - \frac{g}{(2 \pi \hbar)^2}\frac{e^3 f_{0}(0)}{q^2 v_0} I_{\phi}(a,b)}
\end{equation}
Where:
\begin{equation}\label{angle_int}
   I_{\phi}(a,b)=\int\limits_{0}^{2\pi} \Bigg\{ 
     \frac{\cos{\phi}(\cos^2{\phi}-1/2)}{(\cos{\phi}-a)^2+b^2} + 
     \frac{\cos^2{\phi}\sin^2{\phi}(\cos{\phi}-a)}{\big[(\cos{\phi}-a)^2+b^2\big]^2}\Bigg\}d\phi
\end{equation}
with $a = \omega/(q v_0)$ and $b = 1/(q v_0 \tau)$ and that can be analitically calculated from the theory of resudues. We need replace a variable to complex units:

\begin{equation}\label{complex_units}
    \begin{gathered}
         z = e^{i\phi}\\
        \cos{\phi} = \frac{1+z^2}{2z} \\
        d\phi = \frac{1}{i z}dz \\
    \end{gathered}
\end{equation}
Inside of circle with radius 1 we have 3 poles:

\begin{equation} \label{poles}
    \begin{gathered}
    z = a - i b - \sqrt{(a-ib)^2-1} \;\;(second\: order)\\
    z = a + i b - \sqrt{(a+ib)^2-1}\;\; (second\: order)\\
    z = 0 \;\;(first \:order)
    \end{gathered}
\end{equation}

Thus, we can find residues of sub-integral function and find exact form for the integral \eqref{angle_int}.

\begin{equation}\label{angle_int_calc}
   \boxed{I_{\phi}(a,b)= 2 \pi i\Big( ia + \frac{(a- i b)\sqrt{(a-ib)^2-1}}{4b} -  \frac{(a+ i b)\sqrt{(a+ib)^2-1}}{4b} \Big)}
\end{equation}

It is remarkable that second and third terms are complex conjugates, so the sum of them is imaginary. After multiplying the entire bracket by $2\pi i$, it become real. The integral is real for all $a$ and $b$.

The expression \eqref{angle_int_calc} can be interpreted as real function by allocation of the main or secondary branch of the square root of complex function inside. It leads to:

\begin{equation}\label{bad_int}
    \begin{gathered}
        I_{\phi}(a,b)= 2\pi\Big[-a + \frac{\sqrt{|z|}}{2} \Big( \frac{a}{b} \sin{\frac{\theta}{2}} + \cos{\frac{\theta}{2}}\Big) \Big]\\
        \theta = \arccos{\Big(\frac{a^2-b^2-1}{|z|}\Big)}\\
        z = (a + ib)^2 - 1
    \end{gathered}
\end{equation}

It is worth noting that the value range of subcortical function $z$ \eqref{complex_units} lies in the upper part of the complex plane. For this reason we use the function "$\arccos$" and the expression \eqref{bad_int} turns out to be correctly defined for all values of $a$ and $b$.

The system \eqref{bad_int} is resource-intensive for calculating in computer programs, so we are are recommend to use the equation \eqref{angle_int_calc}, besides it contains a multi-valued complex function, that does not affect the computation.

In collisionless electron plasma ($b \rightarrow 0$ and therefore $\tau \rightarrow \infty$) the expression \eqref{angle_int}. leads to:

\begin{equation}\label{angle_intp}
    \boxed{
   I_{\phi}^{(approximate)}(a)= 2 \pi \Bigg( \frac{2a^2-1}{2\sqrt{a^2-1}} -a \Bigg)}
\end{equation}

So, the current \eqref{j2} leads to the expession of $\sigma^{(2)}$ with angle integral that is easy calculated from \eqref{angle_int} for all $a$ and $b$:

The nonlinear conductivity \eqref{sigma2_appendix} in the approximation of fast waves ($q \rightarrow 0$ and low damping ($\omega \tau \rightarrow \infty $) leads to next expresson:

\begin{equation}\label{sigma2_lowq}
    \sigma^{(2)} = - \frac{g}{(2 \pi \hbar)^2}\frac{e^3 f_{0}(0)}{v_0^2} 2\pi \frac{\omega}{q^3}
\end{equation}






    


\end{widetext}

\end{document}